\documentclass[12pt]{article}
\usepackage{amsmath,amssymb,graphicx,mathrsfs,hyperref,slashed,setspace}

\usepackage{float}
\usepackage{latexsym}
\usepackage{bm}
\usepackage{blindtext}
\hypersetup{
	colorlinks=true,
	linkcolor=blue,
	filecolor=black,
	urlcolor=black,
	citecolor=blue,
}

\newcommand{\hide}[1]{}

\renewcommand{\baselinestretch}{1.5}

\newcommand{\be}{\begin{equation}}
\newcommand{\ee}{\end{equation}}
\newcommand{\bea}{\begin{eqnarray}}
\newcommand{\eea}{\end{eqnarray}}

\def\({\left(} \def\){\right)}

\begin{document}
\title{\vspace{-1.8in}
{Thermodynamics of  frozen stars}}
\author{\large Ram Brustein${}^{(1)}$,  A.J.M. Medved${}^{(2,3)}$,
Tamar Simhon${}^{(1)}$
\\
\vspace{-.5in} \hspace{-1.5in} \vbox{
\begin{flushleft}
 $^{\textrm{\normalsize
(1)\ Department of Physics, Ben-Gurion University,
   Beer-Sheva 84105, Israel}}$
$^{\textrm{\normalsize (2)\ Department of Physics \& Electronics, Rhodes University,
 Grahamstown 6140, South Africa}}$
$^{\textrm{\normalsize (3)\ National Institute for Theoretical Physics (NITheP), Western Cape 7602,
South Africa}}$
\\ \small \hspace{0.57in}
   ramyb@bgu.ac.il,\  j.medved@ru.ac.za,simhot@post.bgu.ac.il
\end{flushleft}
}}
\date{}
\maketitle

\renewcommand{\baselinestretch}{1.15}
\begin{abstract}
The frozen star is  a recent proposal for a non-singular solution of Einstein's equations that describes an  ultracompact object which closely resembles a black hole from an external perspective. The frozen star is also meant to be an alternative, classical description of an earlier proposal, the highly quantum polymer model.  Here, we show that the thermodynamic properties of frozen stars closely resemble those of black holes: Frozen stars radiate thermally, with a temperature and an entropy that are perturbatively close to those of black holes of the same mass.  Their entropy is calculated using the Euclidean-action method of Gibbons and Hawking.   We then discuss their dynamical formation by estimating the  probability for a collapsing shell of ``normal" matter to transition, quantum mechanically, into a frozen star. This calculation followed from a reinterpretation of a transitional region between the Euclidean frozen star and its Schwarzschild exterior as a Euclidean instanton that mediates a phase transition from the Minkowski interior of an incipient Schwarzschild  black hole to a microstate of the frozen star interior. It is shown that, up to negligible corrections, the probability of this  transition  is $e^{-A/4}$, with $A$ being the star's surface area. Taking into account that the dimension of the phase space is $e^{+A/4}$, we conclude that  the total  probability for the formation of the frozen star is of order unity. The duration of this transition is estimated, which we then use to argue, relying on an analogy to previous results,  about the scaling of the magnitude of the off-diagonal corrections to the number operator for the Hawking-like particles. Such scaling was shown to imply that the corresponding Page curve indeed starts to go down at about the Page time, as required by unitarity.
\end{abstract}
\maketitle

\newpage
\renewcommand{\baselinestretch}{1.5}
\section{Introduction}

The frozen star was originally meant to provide a classical, geometric alternative to  the collapsed polymer model \cite{strungout}, which describes a highly excited configuration of long, closed, fundamental strings that is lacking  a semiclassical geometric description.  The polymer model is  premised on the notion that the interior of any regular object which successfully mimics a
black hole (BH) should be in a strongly non-classical state \cite{inny}, as follows from the idea that the uncertainty principle prevents the collapse of matter to a singularity, much in the same way that the quantum hydrogen atom is stable against collapse. A strongly non-classical state is tantamount to having maximal entropy density,  which in turn means maximally positive radial  pressure for a state of fixed energy density. The frozen star model aims to mimic the properties of this highly  quantum state in terms of a classical geometry, which amounts to ``flipping'' the radial pressure from maximally positive to maximally negative \cite{bookdill,BHfollies}. The name ``frozen star'' was first adopted  in a later article \cite{popstar} as an homage to classic literature \cite{RW}. The current version of the frozen star model, along with many technical details, can be found in \cite{trajectory} (see also \cite{fluctuations}).

The relevant aspects of the model will be reviewed in the next section, but let us take immediate note of some of the important features of the frozen star. It has been known for quite a while that large negative pressure provides a way  for stabilizing ultracompact objects against further collapse \cite{Buchdahl,chand1,chand2,bondi} and  for evading singularity theorems \cite{PenHawk1,PenHawk2}. Lesser known is that a star with maximally negative radial pressure will be  ultrastable against perturbations \cite{bookdill,popstar,fluctuations}. Another feature is that the modifications from general relativity extend over horizon-sized scales. This happens to be essential for the self-consistency of the model
once  quantum-induced  evaporation is included; otherwise, the radiated energy will far exceed the original mass of the object \cite{frolov,visser}.

As much of the  focus of the current paper is on the thermodynamics of a frozen star, let us recall some of the lore pertaining to ``conventional'' BHs.
Bekenstein \cite{Bek} first proposed that   BHs possess an entropy $S_{BH}$ that is proportional to the surface area of the horizon $A$, and Hawking \cite{Haw}  later
fixed the constant of proportionality according to $\;S_{BH}=\frac{A}{4}\;$. The latter contribution came about when Hawking discovered that a BH of mass $M$, emits thermal radiation with a  temperature of $\;T_H=\frac{1}{8\pi M}\;$. Hawking was able to subsequently  show that the density matrix of the emitted radiation is exactly thermal, leading to what is famously known as the information-loss problem \cite{info}. Closely related to this paradox
is the no-hair theorem \cite{NHbek}, as well as the  transPlanckian problem \cite{trans1}, and less closely related is the BH species problem \cite{species}.
Although the mathematics is consensually  beyond dispute (however, see \cite{helfer}),
interpreting the thermodynamic properties of BHs and resolving the associated  problems remains a challenge in spite of intense efforts.

As the BH entropy is a critical element of the thermodynamic paradigm,   any viable  BH mimicker should be able to account for its exceptionally  large value. We will be able to show that the frozen star can do just that! This is a highly non-trivial challenge, as  even the most compact of stars, when composed of conventional matter, would have an associated  entropy $S$ that is many orders of magnitude smaller than that determined by the Bekenstein--Hawking area law. Indeed, out of the many proposals for ultracompact, almost-black  objects  (See, for example, \cite{carded}), none of these, as far as we know, can replicate  the entropy--area law --- {\em including the factor of 1/4} --- in an unambiguous way.  To understand the difficulty, suppose that we replace the infinite redshift of the BH horizon with a finite but large redshift  $1/\varepsilon^2$,  then what is the line of demarcation between a(n almost) BH and a very compact star? Meanwhile, the BH entropy can be viewed as  a geometric construct that is associated  with (perfectly black) horizons and cannot be assigned generically to regular stars, no matter how compact they might be. This argument is presented  at the end of Gibbons and Hawking's paper  on  BH thermodynamics via Euclidean actions and partition functions \cite{GH}. As we will show, Gibbons and Hawking made a couple of strong assumptions regarding the composition of the stellar material that are not valid for the frozen star; thus our results are not in contradiction with theirs.

An outstanding question about the frozen star is how a collapsing shell of matter might dynamically transition into a  geometry that deviates so dramatically from the Schwarzschild geometry over a horizon-sized length scale.  If this transition does transpire, such a scenario is rather suggestive of a quantum-induced phase transition. Here, we take a first step in support of this idea by applying Mathur's argument \cite{mathur} that such a  transition is  indeed possible  because the large entropy means there is an exponentially large number of microstates  associated with the final-state configuration. And so we  evaluate the probability of transition $\Gamma$, from the Minkowski interior of collapsing shell of matter, or incipient BH, geometry to that of a frozen star  and then, following Mathur, view this as the quantum-transition probability from the initial state to a single microstate of the frozen star. Our calculation requires us to interpret a transitional layer straddling the  the outer surface of the star  as a Euclidean instanton  that mediates the stated transition.  We find that the action of this instanton is equal to the BH entropy up to perturbatively small corrections  and, by the above reasoning, conclude that the probability to transition to  a single microstate is $\;\Gamma \sim e^{-S_{BH}}\;$.  Finally, applying Mathur's argument about multiplying the single-state probability by the large degeneracy of states, we are then led to a total probability of  order unity, suggesting that such a transition is not only possible but likely.
To our knowledge, this is the first time such an estimate of the transition probability has been successfully carried out for an ultracompact, horizonless object.

The subsequent sections are organized as follows: The  next section provides a very brief review on the frozen star. Longer discussions on its conceptual origins  and  important physical features can be found in \cite{bookdill,BHfollies,popstar,trajectory,fluctuations}. In Section~3, we explain why the temperature of emitted radiation from the star would be perturbatively close to the Hawking value \cite{Haw}.  Included is  a heuristic calculation that is premised on the Schwinger effect \cite{Schwinger}, which was originally presented in \cite{Schwing}. See \cite{TulipFever} for a recent discussion on the temperature of compact objects that is similarly based on the Schwinger effect. Section~4 begins with a general discussion on Euclidean BH thermodynamics, following the Euclidean-action method of Gibbons and Hawking \cite{GH}. We go on to  confirm that ,at leading order, $\;S=A/4\;$  for the frozen star by adopting their  method, albeit in an unorthodox way as the outer surface of the star is no longer  a boundary for  the  Euclidean spacetime like it is in the Schwarzschild  case. This calculation of the entropy is novel and one of the  main results in the paper. We continue  on, in Section~5, to reinterpret a narrow interpolating  layer --- smoothly connecting the  Schwarzschild exterior to the frozen star interior --- as a Euclidean instanton which facilitates the transition from the Minkowski interior of an incipient BH to a  single microstate of the frozen star geometry. We are able to confirm that the probability of this transition is precisely $e^{-A/4}$, at leading order in our perturbative parameters, just as advertised. The duration of this transition is also estimated, and this result is used in Section~6 to argue that the number operator and thus also the density matrix for the frozen star radiation could have non-trivial off-diagonal elements. We further argue, following an earlier treatment \cite{density}, that the magnitude of the off-diagonal elements scales so as to reproduce  Page's famous curve describing a unitary evaporation process \cite{page}.
Finally, a brief overview is presented in Section~7. There, we point out that the
Euclidean picture of the frozen star closely resembles a BH sourced by  a condensate
of stringy winding modes \cite{puncture} (see also, \cite{Giveon:2013hsa}) and remark on the significance of this ``coincidence''.

\section{The frozen star geometry}

Here, we will be reviewing the relevant aspects of the frozen star geometry, especially as it pertains to the following analysis. See \cite{popstar,trajectory} for formal details.

Let us first discuss the frozen star metric. We assume that it is static and spherically symmetric, $\;ds^2 =f(r) dt^2 +\frac{dr^2}{f(r)}+r^2d\Omega^2\;$,
where  the equality $\;g_{tt}=-g^{rr}\;$ follows from the star's characteristic equation of state $\;\rho+p_r=0$\;.

As common to  textbook  discussions, one defines a mass function,
\be
m(r)\;=\;4\pi\int\limits_0^r dx\, x^2 \rho(x)\;\;\; {\rm for} \;\;\; r\leq R\;,
\label{five}
\ee
leading to a  Schwarzschild-like form for the $t$--$r$ sector,
\be
 f(r)\;=\;  1- \frac{2 m(r)}{r}\;.
\label{six}
\ee
If $R$ is the radial size of the star, as implied in Eq.~(\ref{five}), then the total mass is defined as
$\;M=m(R)\;$.

We  now deviate from Schwarzschild by setting $\;f(r)=\varepsilon^2\;$, leading to the following line element,
\be
ds^2= -\varepsilon^2 dt^2+ \frac{1}{\varepsilon^2} dr^2+ r^2d\Omega^2.
\label{FSmetric}
\ee
The dimensionless parameter $\varepsilon^2$ should be regarded as a very small number~\footnote{What we now call $\varepsilon^2$ had, until recently, been called $\varepsilon$. The change was made to emphasize its positivity.} (recently estimated to be smaller than $10^{-22}$  \cite{clowncar}) but always finite. With this choice, the mass function is
\be
m(r)\;=\frac{r}{2} (1-\varepsilon^2)\;,
\label{mofr}
\ee
and so
\be
f(r) \;= \; \varepsilon^2\;.
\label{frnew}
\ee

To understand the significance of the perturbative parameter $\varepsilon^2$, it is useful to consider the limit $\;\varepsilon^2\to 0\;$. In this limiting case, $\;g_{rr} \to \infty\;$ at the surface of the star, so that  we can  identify this limit with the classical one $\;l_P\to 0\;$ ($l_P$ is
the Planck length).
In other words, $\;\varepsilon^2 \sim \frac{l_P}{R}\;$.

The matter densities, or diagonal stress-tensor components, then take the forms
\bea
\label{polyrho}
8\pi   \rho &=& \frac{1-(rf)'}{r^2} \;=\;  \frac{1-\varepsilon^2}{r^2}\;, \\
\label{polypr}
8\pi   p_r &=& -\frac{1-(rf)'}{r^2} \;=\; -\frac{1-\varepsilon^2}{r^2}\;, \\
\label{polypt}
8\pi   p_\perp &=& \frac{(rf)''}{2r}\;=\; 0\;,
\eea
where a prime represents a radial differentiation and note that all of the off-diagonal components are vanishing.

\begin{figure}[t]
\vspace{-.75in}
	\includegraphics[height=8cm]{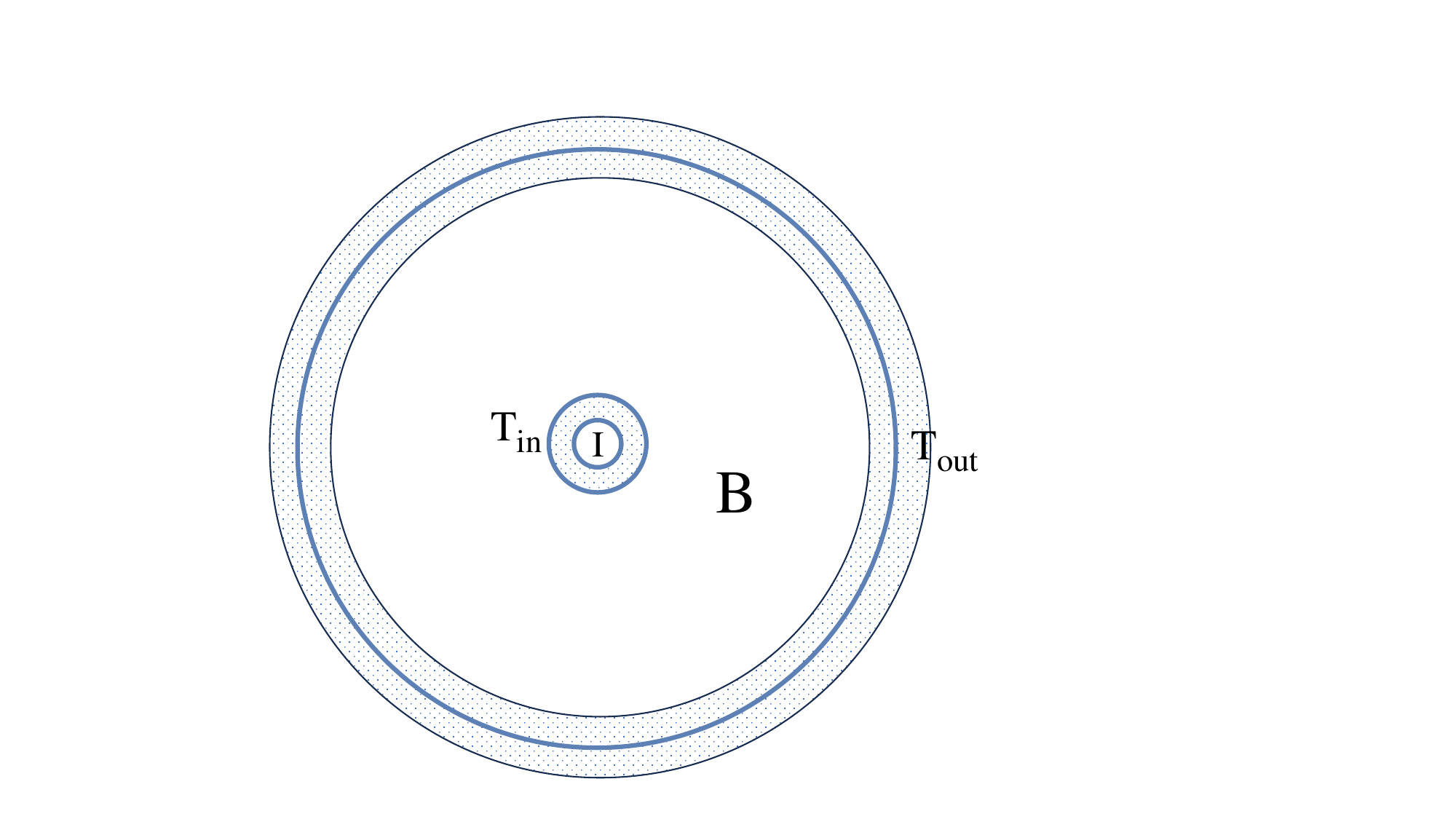}
	\centering
\vspace{-.5in}
	\caption{The frozen star geometry. The core $r\leq\eta R$
		is denoted by $I$, the transitional layer $\eta R<r<2\eta R$ by $T_{in}$, the frozen star bulk $2\eta R\leq r \leq R(1-\lambda)$ by $B$ and the outer transitional layer connecting the star to the Schwarzschild region $R(1-\lambda)\leq r\leq R(1+\lambda)$  by $T_{out}$. Both transitional layers and the core are not drawn to scale, so they can be distinguished in the figure. The actual relative scale is smaller by many orders of magnitude.}\label{fig}
\end{figure}

To ensure that the metric of the frozen star interior joins smoothly to the external Schwarzschild metric, we have added an outer layer $T_{out}$ of width $2\lambda R$, with $\;\lambda\ll 1\;$, that allows for a continuous transition from one geometry to the other \cite{popstar}.    As the density in the bulk of the frozen star  goes as $\;\sim \frac{1}{r^2}\;$ and so is formally divergent at the star's center, there is also a need to smooth out the density in the central region \cite{popstar}. We have
accomplished this  by adding a second transitional layer $T_{in}$ of width $\eta R$ that connects the bulk geometry to a regularized  core of width $\eta R$, where $\;\eta\ll 1\;$. The geometry of the frozen star is depicted in  Fig.~\ref{fig}.

To understand the significance of the perturbative parameter $\lambda$, we call upon  the   polymer model and its string-theoretical perspective. In this picture, the string energy density is decreasing in the transitional  layer from its finite value in the bulk to zero at the
outer surface. This transitional layer needs to still be introduced in the limit $\;\varepsilon^2\to 0\;$, indicating that
$\lambda$ and $\varepsilon^2$ are independent perturbative parameters. Since we have already identified  small values of  $\varepsilon^2$ with the classical limit, it follows that $\lambda$ controls the strength of  classical stringy effects; that is, the effects of  $\alpha'$ or the inverse of the string tension. It can then be concluded that  $\;\lambda^2 \sim \frac{\alpha'}{R^2}\;$.

The non-regularized density leads to small corrections of order $\eta$, when integrated over the entirety of this central region; meaning that the distinction between the bulk and regularized expressions is irrelevant to our leading-order calculations in Section~4. Further note that, in both layers, the condition $\;\rho+p_r=0\;$  is maintained and so it is true throughout the entirety of the star's interior.  As an immediate consequence, the stress-tensor conservation equation reduces from its general static form
 $\; p'_r+ \frac{1}{2}(\ln{f})'\left(\rho+p_r\right)+\frac{2}{r}\left(p_r- p_{\perp}\right)=0\;$ to
\be
p'_r  \;+ \; \frac{2}{r}\left(p_r- p_{\perp}\right)\;=\;0\,
\label{conserv}
\ee
or, equivalently,
\be
p_{\perp}\; = \;\frac{1}{2 r} \partial_r(r^2 p_r)\;.
\label{conserv2}
\ee

We define the interior bulk of the frozen star as the region that ends at $\;r= R\left(1-\lambda\right)\approx 2M\left(1-\lambda+\varepsilon^2\right)\;$, as can be verified by evaluating $\;M=m(R)\;$ via Eqs.~(\ref{five}) and~(\ref{polyrho}). The exterior geometry formally begins at $\;r= R\left(1+\lambda\right) \approx 2M\left(1+\lambda+\varepsilon^2\right)\;$. The outer transitional layer is interpolating between the exterior geometry and the interior bulk.

For future reference, one can view the leading-order scaling relation $\;m=\frac{r}{2}\;$ as meaning
that each spherical slice of the interior is, up to small corrections, like a BH horizon.

\section{Hawking radiation and temperature}

The thermal distribution of outgoing particle emissions from a BH and their associated temperature \cite{Haw}  is traditionally calculated under two assumptions: (1) A future, eternal  event horizon
forms and (2) the exterior geometry can be treated as static. However, it turns out that these assumptions can be relaxed; in which case, compact objects emit Hawking-like thermal radiation provided that  (a) they are compact enough so that the redshift near their surface is sufficiently large  or, equivalently, the gravitational force  $\;F_{\rm G}=\frac{G_N M E}{r^2}\;$ near their surface is strong enough to produce particles with energy $\;E\sim 1/M\;$ in Planck units and (b) a suitable adiabatic condition is satisfied.
These properties were formally defined in \cite{adiabatic}, which stated  that horizonless objects  will indeed emit thermal radiation  but  with two provisions: ($i$) There is  an approximately exponential relation between the affine parameters for the null generators on past and future null infinity
and ($ii$) the surface gravity of the object changes at a slow-enough rate. The frozen star obeys both requirements and is, therefore, expected to emit radiation with properties that are similar to those of BH-emitted Hawking radiation. In the following, we first recall a heuristic argument from \cite{Schwing} (see also \cite{TulipFever}) that relies on a gravitational Schwinger process and then apply the more formal ideas of \cite{adiabatic} to the frozen star.

\subsection{Gravitational particle production analogue of the Schwinger effect}

Let us first consider a pair-production event in the atmosphere of a massive, compact, horizonless object, which is induced by the  gravitational analogue of the  Schwinger effect \cite{Schwing}.  To recall, the Schwinger  equation \cite{Schwinger}  predicts the rate per unit volume ${\cal R}_{PP}^{\rm E\ell}$ of electron--positron pair production in an electric field ${\cal E}$, such that  $\;{\cal R}^{\rm E\ell}_{PP}= \frac{\alpha^2}{\pi^2 }{\cal E}^2  e^{-\frac{\pi m_e^2}{e {\cal E}}}\;$, where $\;\alpha=e^2/(4 \pi)\;$, $e$ is the electron's charge and $m_e$ is its mass. Now suppose that  one exchanges the electric force $\;F_{\rm E}={\cal E}e\;$ with the gravitational force $F_{\rm G}$ in Schwinger's equation. If $M$ is the mass of the compact object and $E$ is the relativistic energy of the positive-energy particle, the  gravitational force is approximately given by $\;F_{\rm G}=\frac{G_N M E}{r^2}\;$, where we restored the dependence on Newton's constant $G_N$. If the object is compact enough so that its radius satisfies $\;R\approx 2M G_N\;$, then $\;F_G \sim\frac{1}{2} \frac{E}{R}\;$. Substituting  $\;m_e^2 \to E^2\;$ and $\;e{\cal E}\to \frac{1}{2} \frac{E}{R}\;$ into ${\cal R}_{PP}^{\rm E\ell}$,
one finds that  the resulting rate of the  gravitational pair production per unit volume ${\cal R}^{\rm G}_{PP}$ in
the proximity of the object's outer surface is
\be
{\cal R}_{PP}^{\rm G}\;\sim\;  \frac{\hbar}{R^4}\left(\frac{R E}{2\pi\hbar}\right)^2  e^{-\frac{2\pi R E}{\hbar}} \;.
\label{heur}
\ee

This rate is maximized when $\;E = \frac{\hbar}{\pi R}\;$, which is of the order of the Hawking temperature $\;T_H = \frac{\hbar}{4\pi R}\;$ of a Schwarzschild BH; in which case, $\;{\cal R}^{\rm G}_{PP}\sim\frac{\hbar}{R^4}\;$. This estimate suggests that one Hawking-like pair of particles, each with energy $\;E\sim T_H\;$,  is produced  per light-crossing time $R$ from a volume $R^3$, which is the standard Hawking emission result. Meanwhile, away from the outer surface, the rate of pair production  becomes exponentially small; indicating that Hawking-like radiation can originate  near the outer surface
of  a compact object, even if it is not  formally a horizon.

\subsection{Temperature and spectrum of emitted radiation}

\subsubsection{Formal conditions for the validity of the calculation}

For a more formal derivation, we recall the results of \cite{adiabatic}, which reveals that horizonless objects  will indeed emit thermal radiation under certain conditions.

Following \cite{adiabatic}, let us first consider null curves starting from past infinity $\mathcal{J}^{-}$ and
arriving at future infinity $\mathcal{J}^{+}$. There will be some invertible relation between their respective affine parameters $U$ and $u$; namely,
$\;U= p(u)\;$ and $\;u=p^{-1}(U)\;$.
The surface gravity can then be defined formally as
$
	\;\kappa(u)=-\dfrac{\ddot{p}(u)}{\dot{p}(u)}\;,
$
where dots denote  derivatives with respect to $u$.
Next,  picking one particular null curve and labelling it by $U_*$ on $\mathcal{J}^{-}$ and $u_*$ on $\mathcal{J}^{+}$,
 one can then  write
\begin{equation}
	U\;=\;U_*+C_*\int_{u_*}^{u}e^{-\int_{u_*}^{\bar{u}}\kappa(\tilde{u})d\tilde{u}}d\bar{u}\;,
\label{big-U}
\end{equation}
for some dimensionless  constant $C_*$.

Assuming that $\kappa(u)$ satisfies the following adiabatic condition :
\begin{equation}
	\frac{\left|  \dot{\kappa}(u_*) \right|}{\kappa^2(u_*)} \;\ll \; 1
\label{adiabatic-con}\;,
\end{equation}
one can further show that outgoing particles reaching $\mathcal{J}^{+}$ at $u_*$ obey a Planckian distribution. In  which case, Hawking-like radiation exists with a temperature of
\begin{equation}
	 T(u_*)\;=\;\frac{\kappa(u_*)}{2\pi}\;.
\end{equation}

These ideas can be  applied explicitly to the frozen star because they rely solely on the nature of spacetime outside the star.
Importantly, the exterior region for $\;r>R_{star}\equiv R\left(1+\lambda\right)\approx 2M\left(1+\lambda+\varepsilon^2\right)\;$ is described by the Schwarzschild metric; thus the affine parameters are the Eddington--Finkelstein outgoing coordinate $\;u=t-r_*\;$, where $r_*$ is the usual tortoise coordinate that  is defined by
$\;dr_*= \frac{dr}{\sqrt{-g_{tt}g^{rr}}}\;$,  and $U$ is a Kruskal-like coordinate that is related to $u$ in the standard way.
One can write $r_*$ for both the interior and exterior regions
and, since $r^*$ is continuous across the inner and  outer surfaces of the translational layer (and throughout the layer itself),  $u$ and $U$ are continuous as well.

The surface gravity for the Schwarzschild metric, when defined at the surface of a star of radius $R_{star}$, is
\begin{equation}
	\kappa^2\;=\;-\frac{1}{2}\left( \nabla_\mu \xi_\nu \right)\left(\nabla^\mu \xi^\nu \right)|_{r=R_{star}}\;,
\label{kappahouse}
\end{equation}
where $\xi$ is the timelike Killing vector. As in the  case of a Schwarzschild BH, $\kappa$ can be expressed as $\;\kappa(r)=\frac{M}{r^2}\;$, so that, on the surface  $\;r=R_{star}\;$ of a frozen star,  the leading-order  result is simply   $\;\kappa=\frac{M}{R^2}\;$.

With the above relations in hand, we can  now define the affine parameters $U$ on $\mathcal{J}^{-}$ and $u$ on $\mathcal{J}^{+}$
for the frozen star,
\begin{equation}
	U\;\approx\; U^*_H-A_*e^{-\kappa_* u}\;,
\end{equation}
\begin{equation}
	u\;=\;-\frac{1}{\kappa_*}\ln\left(\frac{U^*_H-U}{A_*}  \right)\;,
\end{equation}
where $A_{*}$ is a dimensional constant,
the best estimate for the would-be horizon is  at $\;U^*_H=R_{star}\approx R\;$ and
notice that $\kappa_*$ satisfies $\;\kappa_*=-\ddot{U}/\dot{U}\;$.

The surface gravity  naively  vanishes in the interior of the frozen star because of the constancy of
$g_{tt}$ and $g^{rr}$. However, as we will argue  later,  each spherical slice of
the interior maintains the same properties that one would attribute to a would-be horizon
of that radial size. The physical reason that these inner slices do not directly contribute to the radiation is not that their Killing vector vanishes but
because of the large interior redshift  greatly suppressing the propagation of particles
in the bulk of the frozen star. So that, even if such particles are produced, they are effectively trapped within the star.

The adiabatic condition~(\ref{adiabatic-con}) for the frozen star model can be verified by
by evaluating the derivatives of the  surface gravity. In general, for a star of radius $R_{star}$,
\begin{equation}
	\dot{\kappa}\;=\;\frac{d\kappa}{du}\;=\;\frac{\partial\kappa}{\partial t}-f(r)\frac{\partial\kappa}{\partial r}
|_{r=R_{star}}\;.
\end{equation}
In the frozen star case, $\;\frac{\partial\kappa}{\partial t}=0\;$ and $\;\kappa(r)=\frac{M}{r^2}\;$, and so
\begin{equation}
	\dot{\kappa}\;=\;\frac{2M}{r^3}f(r)|_{r=R_{star}}\;,
\label{kap-dot}
\end{equation}
which at leading order becomes
\begin{equation}
	\dot{\kappa}\;=\;\frac{\varepsilon^2+\lambda}{ R^2}\;.
\end{equation}
On the other hand, at $\;r=R_{star}\;$, another leading-order result is
\begin{equation}
	\kappa^2\;=\;\frac{1}{4R^2} \;,
\end{equation}
so that the adiabatic condition~(\ref{adiabatic-con}) translates into
\begin{equation}
	\frac{|\dot{\kappa}|}{\kappa^2}\;=\; \frac{1}{4}\left(\varepsilon^2+\lambda\right)\;\ll\; 1\;,
\label{adia-con}
\end{equation}
meaning that the adiabatic condition for the frozen star is easily satisfied.

\subsubsection{Hawking-like radiation from a frozen star}

We are now well equipped to derive the thermal spectrum of the emitted radiation for a frozen star by
suitably adapting the derivation in \cite{adiabatic}.

Defining the modes on $\mathcal{J}^{-}$ as
\begin{equation}
	\xi(u)\;=\;\frac{1}{\sqrt{4\pi \omega}}e^{-i\omega u}
\end{equation}
and those on $\mathcal{J}^{+}$ as
\begin{equation}
	\varPhi(U)\;=\;\frac{1}{\sqrt{4\pi\Omega}}e^{-i\Omega U}\;,
\end{equation}
we can express the Bogoliubov coefficient
\be
 \alpha_{\omega \Omega}\;=\; i\int dU\left[ \partial_{U}\varPhi^*\xi-\varPhi^*\partial_{U} \xi  \right]\;
\ee
as
\be
	\alpha_{\omega \Omega}\;=\;-\frac{1}{4\pi\sqrt{\omega\Omega}}\int_{-\infty}^{U^*_H} dU e^{i\Omega U}\left[\Omega\left(\frac{U^*_H-U}{A_*}\right)^{\frac{i\omega}{\kappa_*}}+\frac{\omega}{\kappa_*}\left(\frac{U^*_H-U}{A_*}\right)^{\frac{i\omega}{\kappa_*}-1}
	\right]\;.
\ee

Recalling that $\;\beta_{\omega \Omega}\approx -i \alpha_{\omega \Omega}\;$ and $\;i=e^{i\frac{\pi}{2}}\;$, we
then find that
\begin{equation}
	\beta_{\omega \Omega}\;=\;-i\frac{1}{2\pi}\sqrt{\frac{\omega}{\Omega}}\frac{(\Omega)^{-\frac{i\omega}{\kappa_*}}}{\kappa_*} e^{\frac{\pi \omega}{2\kappa_*}}\Gamma\left( \frac{i\omega}{\kappa_*} \right)\;,
\end{equation}
which  allows us to calculate
the expectation value of the number operator for the outgoing particles via the usual definition
$\;\langle 0| b^\dagger b |0\rangle \;=\;\int_0^\infty d\Omega \beta_{\omega \Omega}\beta^*_{\omega' \Omega}\;$. The result is
\begin{equation}
	|\beta_{\omega \Omega}|^2=\frac{1}{e^{\frac{2\pi\omega}{\kappa_*}}-1}\delta(\omega-\Omega)\;.
\end{equation}

The frozen star temperature can now be readily identified as
\begin{equation}
	T_{FS}\;=\;\frac{\kappa_*}{2\pi}\;=\;\frac{1}{8 \pi M }\;=\;T_H\;,
\end{equation}
to leading order in $\varepsilon^2$ and $\lambda$.

In summary, the temperature of the emitted radiation from a frozen star is the same as Hawking temperature up to a perturbatively small correction.



\section{Entropy}

We start  here with a  calculation of  the entropy of the frozen star by integrating the first law, as was first done by Bekenstein and Hawking \cite{Bek,Haw}. Following this, the  entropy of the frozen star will be calculated using the Euclidean-action method as pioneered  by Gibbons and Hawking \cite{GH}.

\subsection{Entropy from the first law}

The first law states, in the current context, that
\be
dS\;=\;\beta(M) dM \;,
\ee
where $\beta(M)$ is the inverse of the temperature.

To derive the entropy of a frozen star of mass $M$, it is helpful to first consider a star of much-smaller mass $\;m_0\ll M\;$. From the calculation of the temperature in the previous section, the inverse temperature at leading order is
\be
\beta(m_0)\;= \;8\pi m_0\;.
\ee

One can now imagine increasing the mass of the frozen star by a small amount $\Delta m$, so  that  $m_0+\Delta m$ is the new mass and  and  $8\pi (m_0+\Delta m)$ is the new inverse temperature. It follows that
\be
S_{FS}(m_0+\Delta m)\;=\; S_{FS}(m_0) + 8\pi m_0 \Delta m +{\cal O}(\Delta m^2)\;.
\ee

Integrating the previous  equation,  with the key observation  that $\beta(m)$  grows linearly with $m$,~\footnote{A similar scaling relation is
used in \cite{Ash} to define an effective surface gravity.}
\be
\beta(m)\;=\; 8\pi m\;,
\label{betam}
\ee
one obtains
\bea
S_{FS}(M)&=&  S_{FS}(m_0) + \int_{m_0}^M  dm~8\pi m
\;=\; S_{FS}(m_0)+ 4\pi \left(M^2-m_0^2\right)\nonumber \\ &=& S_{FS}(m_0)+S_{BH}(M)-S_{BH}(m_0)\;.
\eea

If we start out with the frozen star's seed radius being close to its core radius, then   $S_{FS}(m_0)$ and $S_{BH}(m_0)$ are negligible
in comparison to $S_{BH}(M)$ . The conclusion is that, to leading perturbative order, the frozen star entropy is equal to the Bekenstein--Hawking entropy of a same-mass BH,
\be
S_{FS}(M)\;=\; S_{BH}(M)\;.
\ee

\subsection{Entropy from Euclidean action}

Next is the calculation of the entropy of the frozen star using the method of Gibbons and Hawking \cite{GH}. We start by calculating the action of the bulk of the frozen star, which extends from the center up to the beginning of the outer transitional layer. We then calculate the entropy of the bulk and show that it is equal, at leading order in $\varepsilon^2$ and $\lambda$,  to the entropy of a  same-mass BH.
The action of the outer transitional layer will then be calculated separately. In the next section, this outer layer will be interpreted as a Euclidean instanton that mediates the transition between a collapsing shell of ordinary matter with a Minkowski interior and the bulk of a frozen star.  The current section ends with a calculation of  the  entropy for the total spacetime, which  again agrees at  leading order
with  that of a BH of  the same mass.

The Euclidean-action method then provides us with  two ways of determining the frozen star entropy; either  by calculating the action of the bulk of the interior or by evaluating the action for the total spacetime.   This duality then provides us, in turn, a means for realizing  both of the competing  perspectives on the entropy, matter based and geometric, but with no ambiguity in the final answer.  Note that  the geometric  point of view follows automatically if the limits $\;\varepsilon^2 \to 0\;$ and $\;\lambda \to 0\;$ are imposed; in which case, the outer surface of the star tends to  that of a true horizon.  However, it is more interesting and realistic to consider what happens when $\varepsilon^2$ and $\lambda$ differ from zero.

Let us now recall how the calculation in \cite{GH} is performed.
Gibbons and Hawking started out  by evaluating the action of the Euclidean BH solution $I[g]$ and applying the identification
\be
I[g]\;=\;\ln{\cal Z\left[\beta\right]}\;=\;-\beta F \;=\; S- \beta M \;,
\label{IF}
\ee
where  $\ln{\cal Z\left[\beta\right]}$ is the logarithm of the thermal partition function and       $F$ is the Helmholtz free energy.

The Euclidean Einstein action consists of both a volume and a surface contribution,
\be
I[g]\;=\;-\frac{1}{16\pi}\int_{\cal M} \sqrt{g} {\cal R} \;-\;\frac{1}{8\pi} \int_{\partial{\cal M}} \sqrt{h} [K]\;.
\label{eucac}
\ee
The integrand of the volume portion trivially vanishes for the Schwarzschild solution, and one is left only with the boundary contributions from  the horizon and timelike infinity, each of which is an integral of the trace of the extrinsic curvature $K$  (or second fundamental form), where
$[K]$ indicates that the contribution from flat space has been subtracted off.
The Euclidean geometry has zero area at the horizon and, once the conical singularity at the horizon has been resolved by assigning the Euclidean time $\tau$ with its appropriate periodicity, $\;\Delta \tau =\beta=8\pi  M=4\pi R\;$, the contribution from the horizon vanishes.

The only surviving  contribution is then that of the asymptotic boundary $I_{\infty}$. The Schwarzschild result goes as
\be
  I_{\infty}\;=\;-\frac{1}{8\pi} \oint d\tau \sqrt{g^{rr}} \partial_r\left( 4\pi r^2 \sqrt{g_{\tau\tau}}\right)_{|r\gg 2M}\;.
\ee
After appropriately subtracting the divergent flat-space contribution and taking the $\;r \to\infty\;$ limit at the end, one finds that
\be
  I\;=\; -\frac{1}{2} \beta M\;.
\label{eucinf}
\ee

Now recalling Eq.~(\ref{IF}),
one arrives at
an  entropy of
$\;S= \beta M/2\;$ or, since $\;\beta=8\pi  M\;$, $\;S=4\pi M^2\;$. This is simply $\;S=S_{BH}=A/4\;$ and thus the  Bekenstein--Hawking area law \cite{Bek,Haw}.
For future reference, one should notice that $\;I_{\infty} =-S_{BH}\;$.

The Gibbons--Hawking calculation relies on the fact that the Euclidean action is a boundary term on the solution of the Einstein equations of motion. Clearly, if the sole contribution to the Euclidean action comes from the boundary term at infinity, one is guaranteed to get exactly the BH entropy. In the original calculation, the interior of the BH is excised, the horizon shrinks to a point and the conical singularity is removed. Therefore,  the sole contribution
does indeed come from the asymptotic  boundary term. However, in Section~5 of \cite{Brustein:2022wiq}, it was pointed out that the same result is still obtained for a horizonless and non-singular spherical geometry for which the volume of the sphere at the origin shrinks to zero. In this case, there is no contribution from the inner boundary and the entropy is exactly equal to that of a BH. We will encounter a similar situation in what follows.

\subsubsection{Entropy of the frozen star bulk}

\begin{figure}[h]
\vspace{.5in}
	\includegraphics[height=8cm]{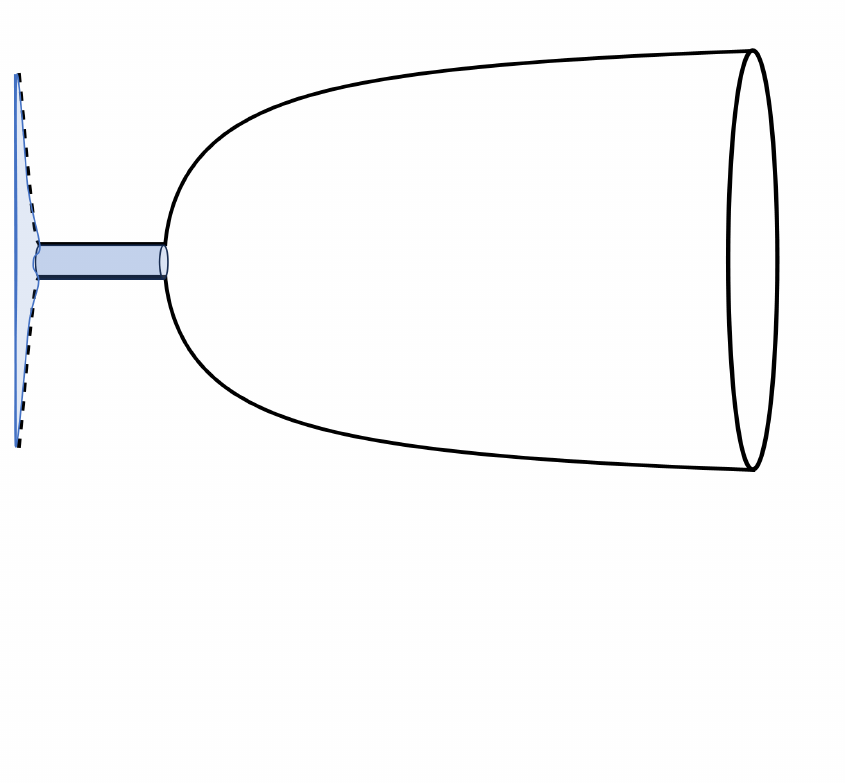}
	\centering
\vspace{-1in}
	\caption{The frozen star Euclidean geometry. The exterior geometry is essentially the same as the Euclidean Schwarzschild geometry. The bulk has the geometry of a cylinder of radius $\sqrt{\varepsilon^2}$ and the geometry of the inner transitional layer, drawn here not to scale, expands back to the asymptotic radius.}
\label{figEuclidean}
\end{figure}

To calculate the entropy of the bulk (again to leading order in $\varepsilon^2$ and $\lambda$), it is important to  realize that $\beta$ depends on $r$. From Eq.~(\ref{betam}) and from the relation $\;m(r)=\frac{1}{2} r\;$ in Eq.~(\ref{mofr}), it follows that
\be
\beta(r)\;=\;4\pi r \;,
\label{betablockers}
\ee
which is obviously different than the constant   $\beta_H = 8 \pi M\;$ in \cite{GH}.
A radially dependent compactification scale may seem unusual, but  then so
is the frozen star, which features large deviations from Schwarzschild over the entirety of its interior. The Euclidean geometry of the frozen star is depicted in Fig. \ref{figEuclidean}. A similar geometry was found in \cite{Giveon:2013hsa}.

It is possible to understand this peculiar dependence by considering the near-horizon geometry of the Euclidean Schwarzschild BH, for which  the relevant part of the metric can be expressed as  $\;\frac{4\pi^2}{\beta_H^2} \tilde{r}^2 d\tau^2 +d\tilde{r}^2\;$. Now, replacing the constant $\beta_H$ by $\;\beta \sim {r}\;$ and  taking into account that $\;\tilde{r} \sim r\;$ for the frozen star interior, one  finds  a constant  $g_{\tau\tau}$.
In any case, such radial dependence inside of the frozen star does not affect the validity of the method. As we have already observed in Section~3, in the exterior of the frozen star, the temperature  is a constant and its inverse is  equal at leading order to the standard value of $\;\beta_H\;$.

The Euclidean action of the bulk  does not receive any contribution from the extrinsic-curvature boundary terms because, as argued above, it vanishes at $\;r=0\;$ due to the  shrinking of the volume of the sphere to zero at the origin and the outermost surface  of the bulk is not a true boundary of spacetime. The remaining volume integral is
\bea
I_{Bulk} &=& -\frac{1}{16\pi}\int_{Bulk} d^4x\;\sqrt{g} \; {\cal R}\;,
\eea
and using  Eq.~(\ref{betablockers}) for  the compactification scale of the Euclidean time direction, one can rewrite this as
\be
I_{Bulk} \;=\;
 -\frac{1}{4} \int_0^R dr \; 4\pi r^2 \;r {\cal R}\;.
\ee

Let us next use the trace of Einstein's equations, while recalling the form of the  stress
tensor  ${\cal T}^{a}_{\;\;b}$ from
Section~2, to recast the scalar curvature in terms of the energy density and pressure  in the frozen star's bulk,  $\;{\cal R} = -8\pi {\cal T} =
-8\pi\left(-\rho+p_r+2p_{\perp}\right)=16\pi \rho=\frac{2}{r^2}\;$.  So that
\be
I_{Bulk} \;=\;
 - 2\pi \int_0^R dr  \; r  =  -\pi R^2  \;
\label{lights,camera}
\ee
and, therefore,
\be
I_{Bulk} \;=\; -\frac{1}{2}\beta(R)M\;.
\ee
Using the relationship~(\ref{IF}) between the action and the entropy, we conclude that $\;S_{Bulk}=S_{BH}\;$.

It is worthwhile to understand how we managed to evade Gibbons and Hawking's conclusion that a normal star or, for that matter, any horizonless object could not have a geometric entropy, regardless of how compact it may be \cite{GH}. Looking at the discussion leading up to their Equation~(3.16),
one can identify two critical assumptions: (1) The star is composed of ``normal" matter in thermal equilibrium and (2) its radius $R$ and temperature $T$  satisfy $\;R^3\gg T^{-3}\;$. Note that $T$ here is not the Hawking temperature but rather the actual thermodynamic temperature of the matter. The former condition does not apply to our model, as the source of the frozen star is exotic matter at zero temperature.~\footnote{Fundamentally,
the frozen star would consist of a state of strings close to  Hagedorn temperature \cite{strungout}, but this matter is no less  exotic.} As for the latter condition, this is clearly violated as it fails whether we use the inverse of the Hawking temperature, which is order $R$, or the actual inverse temperature, which formally diverges.

\subsubsection{Action of the outer transitional layer}

It is straightforward to evaluate the action of the outer transitional layer $I_{TL}$ using the  metric and stress-tensor components as presented in
Section~2. Recall that this transitional layer is a thin shell of width $2\lambda R$, with $\;\lambda\ll 1\;$. The metric of this layer smoothly connects  the bulk metric to the exterior Schwarzschild metric.  For simplicity,  we evaluate the action to leading order in $\varepsilon^2$ and $\lambda$ and similarly  for the rest of the calculations. Here, we only need to evaluate the volume term in the action because the boundary terms are supported only on the true boundaries at $\;r=0\;$ and $\;r\to\infty\;$, which are outside the domain of integration. In this layer, it remains true that $\;p=-\rho\;$, but $p_{\perp}$ is non-vanishing and large, so that now $\;{\cal R}=-8\pi {\cal T}=16 \pi(\rho-p_\perp)\;$. It follows that
\be
I_{TL}  \;=\;
-4\pi \beta \int_{TL}~ dr  \; r^2   (\rho-p_\perp)\;,
\label{ITL}
\ee
where, in principle, $\beta$ could be a function of $r$; however, since the width of the boundary layer scales as $\;\sim \lambda\;$ and the value of $\beta$ at the outer edge of the layer is $\;\beta=4 \pi R\;$,  we can ignore any possible $r$ dependence at leading order.

We now recall our first form of the frozen star conservation  equation~(\ref{conserv})  and  also consider that  $\;p'_r \sim \dfrac{|p_r|}{\lambda}\gg |p_r|,\rho\;$,
from which it can be deduced that $\;p_{\perp} \sim \dfrac{p_r}{\lambda}\gg |p_r|\;,\rho\;$. Hence, we can ignore the contribution from $\rho$ in
Eq.~(\ref{ITL}) at leading order.
Also recalling our second  form of the conservation equation~(\ref{conserv2})
and  observing  that the factors of $r$ appearing  in Eq.~(\ref{ITL}) can be approximated  as $R$'s, we can reexpress the action $I_{TL}$ as
 \be
I_{TL}\;=\; 16\pi^2 R\int\limits_{R(1+\varepsilon^2-\lambda)}^{R(1+\varepsilon^2+\lambda)}  dr
~\frac{R}{2}~ \partial_r(r^2 p)\;=\;
 8\pi^2 R^2~ (r^2 p)\biggl|^{R(1+\varepsilon^2+\lambda)}_{R(1+\varepsilon^2-\lambda)}\;.
\label{ITL1}
\ee

The value of $p$ at the upper end is its Schwarzschild value of $\;p =0\;$, while the value of $p$ at the lower end is its frozen star bulk value
of $\;r^2 p=-\frac{1}{8\pi}\;$, meaning that
\be
I_{TL}\;=\; + \pi R^2 \;=\; S_{BH}\;.
\label{ITL2}
\ee

\subsubsection{Total entropy of the frozen star}

The action of the Schwarzschild exterior is given by
\be
I_{EXT}\;=\;-\frac{1}{16\pi}\int_{r > R(1+\varepsilon^2+\lambda)} dx^4 \sqrt{g} {\cal R}\;\;-\; \frac{1}{8\pi} \int_{\partial {\cal M} }\sqrt{h} [K]\;.
\label{eucext}
\ee
The volume term vanishes for the exterior Schwarzschild metric and the surface contribution comes only from the boundary at  infinity, as the boundary at $\;r=0\;$ is outside the domain of integration.

The contribution of the boundary at infinity is  given in Eq.~(\ref{eucinf}) and is equal to $-S_{BH}$, so that
\be
I_{EXT}\;=\;-\frac{1}{2} \beta M \;=\; -S_{BH}\;,
\label{eucext1}
\ee
which can be combined with our previous calculations to give
\be
I[\text{frozen star}]\;=\;I_{Bulk}+I_{TL}+I_{EXT}\;=\;-S_{BH}+S_{BH}-S_{BH}\;=\; -S_{BH}\;.
\ee
It follows that the frozen star has a leading-order entropy that is equal to that of a  BH with the same mass,
\be
S_{FS}\;=\;S_{BH}\;.
\ee

Let us comment that, because the outer surface becomes a true horizon in the $\;\varepsilon^2 , \lambda\to 0\;$ limits, it
must be true that the zeroth-order contributions to $I_{Bulk}$ and $I_{TL}$ exactly cancel, just as was found above.
This tells us, unequivocally,  that the scaling relation $\;\beta =4\pi r\;$ is the correct choice as anything else
would have offset the required cancellation.

\section{Dynamical collapse of matter to a frozen star}

As suggested  in the Introduction, it is unreasonable to expect a ``normal''  system of collapsing matter  to slowly evolve into  an ultracompact object whose geometry  deviates from the Schwarzschild geometry
on horizon-order length scales. What is rather needed is something akin to a quantum-induced phase transition, where we have implicitly recalled
that the final system of matter  is, in its fundamental description,  a highly quantum state of strings at a finite temperature.
It is then natural, in the Euclidean picture, to regard the transitional  layer as a gravitational instanton that mediates the
transition
from an infalling shell of matter with a Schwarzschild exterior and an empty interior to  a frozen star.

There is, indeed, a  long tradition of using Euclidean instantons in just this way, whether it has been to facilitate bounces between two phases of de Sitter ({\em e.g.}, \cite{coleman}),
to induce a quantum-tunneling process  from nothingness to a Lorentzian inflationary cosmology ({\em e.g.}, \cite{HH}) or, in the guise of a  domain wall, to trigger the creation of a pair of  BHs ({\em e.g.}, \cite{GGS}). For a general discussion, see \cite{BC,Brown:2007sd}.   The theme in all these works, in analogy to similar ideas in quantum mechanics,  is that the probability of transition is given (up to a numerical prefactor) by $\;\Gamma= e^{-I_{Inst}}\;$, where the instanton action $I_{Inst}$ is typically  evaluated using either  the on-shell solution or at a saddle point of an appropriate partition function. The instanton interpolates between an initial spacelike section of some specific  solution, which is typically the false vacuum solution,  and a final spacelike section of another solution,
so that the action $I_{Inst}$ can be defined by a subtraction procedure of the respective Euclidean actions of  these two solutions.

In our case, the initial-state geometry is defined on the outermost slice of the transitional  layer and corresponds to the  solution when the collapsing shell of matter, which can also be viewed as a domain wall,  is just on the brink of passing through its would-be horizon. The interior of the shell at this initial time contains the false vacuum, which is simply flat Minkowski space. All this  is in agreement with our calculation at the end of the last section, which makes it clear that the Euclidean action vanishes on the initial-state spacelike slice, since a Schwarzschild exterior enclosing normal matter will have a negligible free energy until a horizon forms.

Meanwhile, the final-state geometry is defined on the innermost slice of the transitional layer, at a radius just inside the would-be horizon.~\footnote{One of our working assumptions is that $\;\varepsilon^2<\lambda\;$ (see Section~2), and so
$\;R+\varepsilon^2-\lambda<R\;$.} Once this time has been reached, the shell and its once-empty interior has transitioned into a fully formed frozen star. From this perspective, the inner-most slice represents a bubble enclosing the true vacuum, as the frozen star represents a state
of maximal entropy or minimal free energy.

This picture depends on having, from an external perspective,  a   finite initial  thickness for the matter shell and a slow velocity for its descent. Given that the shell's width far exceeds that of the transitional  layer, $2\lambda R$,
and that its collapse is slow, $\;\dfrac{dr}{dt}\sim\lambda\;$,
the phase transition can be triggered long before the entirety of the shell passes through its Schwarzschild radius.

The standard prescription in the case of a bubble  or bounce (B) mediated transition at finite temperature  from a false vacuum (FV) to the true one
is an exponent of the form  \cite{BC,Brown:2007sd}
\be
\ln\Gamma\;=\; I_{B}-I_{FV}\;,
\ee
which is, in our case,
\bea
\ln\Gamma &=& I\left[R(1+\varepsilon^2-\lambda)\right]-I\left[R(1+\varepsilon^2+\lambda)\right]\nonumber \\
&=&  -I_{TL} \;,
\eea
and so
\be
\Gamma \;\sim\; e^{-S_{BH}}\;,
\ee
as previously advertised. Formally, in the classical limit of $\; l_P\to 0\;$, the transition probability vanishes and the frozen star never forms;
 the  shell rather  continues on to form a singular BH. As will soon be shown, there is evidence that, if a frozen star does form, its evaporation is unitary, painting a nicely consistent picture.

We now follow Mathur ({\em e.g.}, \cite{mathur}) and identify this as the probability for transition to any one
of the $\;e^{S_{FS}}=e^{S_{BH}}\;$  frozen star microstates, so that the total probability is
\be
\Gamma_{Tot}\;\sim\; e^{+S_{BH}} \Gamma \;\sim 1\;\;.
\ee
That is to say, the transition  from an incipient BH
to any frozen star configuration of equal mass  is a likely, if not certain, outcome.

\subsection{How fast?}

As for the duration of the transition from an external observer's perspective, this is simply the width of the transitional layer $2\lambda R$, divided by the shell's
approximate radial velocity in the outer layer $\;\lambda\pm\varepsilon^2\sim \lambda\;$. In  other words, a time scale of order $R$, the light-crossing time of the star.  Consider though that the analogue calculation for  a BH  is exponentially large. However, from the perspective of a particle that exits the matter shell just before the frozen star has fully formed, the relevant time scale is the proper-time equivalent or, quite simply, the width of the transitional layer divided by a speed of order unity, $\;T_{trans}\sim\lambda R\;\sim \sqrt{\alpha'} \sim l_s\;$, where
we have recalled a discussion from Section~2 relating $\lambda$  to the inverse of the string tension  and $l_s$ is
the fundamental string length, which in weakly coupled string theory satisfies  $\;l_P= l_s g_s\;$ with $g_s$ being the string-coupling strength.

The scale $T_{trans}$  can  be compared to the analogue time scale in general relativity that  is used in the calculation of the Hawking effect, where the transition (or horizon formation) is assumed to be instantaneous.  This string-scale fuzziness in the transition time --- or, equivalently,  in the  thickness of the transitional layer  --- will be important in what follows.

\section{Radiation revisited}

In this section, we argue on the basis of  previous discussions in \cite{slowleak,density} that, when the entropy $S_{BH}$ is finite, the evaporation of the radiation is consistent with unitarity.  The key observation is that this finiteness implies that the would-be horizon is not a surface of infinite redshift but rather a surface of large, finite redshift and, consequently, has some quantum width. As a result, the number operator
of the Hawking particles is no longer diagonal and  likewise for the density matrix of the emitted radiation.

It proves to be a worthwhile exercise to think about the radiation process of the frozen star in the context of Hawking's seminal calculation of the thermal spectrum \cite{Haw}.
In Hawking's original model, one considers a collapsing shell of matter, viewed as an incipient BH, and focuses on a certain class of null rays. Namely, those rays which enter the interior of the shell when  its  radius  $R_{shell}$ is substantially larger than its Schwarzschild radius $R$ but  exit the interior at a time close to that  of horizon formation, so that $\;(R_{shell}-R)_{in}\gg (R_{shell}-R)_{out}$. Because the ray exits when $\ (R_{shell}-R)_{out}/R\ll 1\;$, it  undergoes an exponentially large redshift after passing through the interior region. From the perspective of an asymptotic observer, it is this strong time-dependent effect that leads to the creation of Hawking particles.  Hawking was able, of course, to make this description of particle emission from
an incipient BH  quantitatively precise.

The analogous situation for the frozen star is that the null ray must enter the  collapsing matter shell before the transition (as described in Section~5) takes place but then exit the shell only after the transition has transpired. This is the only way in which the ray can experience the effect of a suitably large enough redshift. It can then  be deduced that, if the transition occurs at an advanced time of $\;v=v_{trans}\;$,
the prospective Hawking mode should have entered the matter distribution no later
than this time. What should now be clear is that  $v_{trans}$
is the analogue of Hawking's $v_0$, which is defined by the last incoming ray that can  exit before the horizon finally forms. For now on, we will denote both of these cutoff times by $v_{\star}$.

If the transition from ordinary matter to a frozen star was an instantaneous process, this would be the end of the story, as the mathematics would basically replicate those of the Hawking calculation. However, as discussed in the previous section,  the time scale of the transition is, locally, given by $\;T_{trans}\sim \lambda R\sim l_s\;$;~\footnote{We use the local time scale $R\lambda$ and not  the asymptotic scale $R$ as what is required is the analogue of Hawking's instantaneous-collapse model.} meaning that $v_{\star}$ in the frozen star model should not  be regarded as precise but should rather be smeared over this scale. Alternatively, one could view the smearing as an uncertainty in the location of the would-be horizon.
Either way, this smearing modifies the Hawking calculation in a subtle but important manner. Fortunately, we are well positioned  to account for just such an effect, as will now be explained.

Let us recall an earlier discussion \cite{slowleak}, where the effects of a quantum-uncertain horizon position were incorporated into the otherwise standard Hawking calculation.  Here, we will just sketch the argument, leaving a precise calculation to  a future publication.

The key revision of Hawking's calculation in \cite{slowleak} is based on an integral that appears in
Hawking's original calculation of the number operator for the outgoing particles,
\be
I_C\;=\;\frac{1}{2\pi}\int\limits_{-\infty}^{\infty} dv \; e^{i v (\omega'-\omega'')}\;=\;\delta(\omega'-\omega'')\;,
\ee
where $\omega'$ and $\omega''$ are incoming-particle frequencies and the advanced time is defined as $\;v=t+r_*\;$. For a formal definition of the number operator, see Section~3.
This integral is to be adapted so that it includes the effects of semiclassical physics. In \cite{slowleak}, these effects resulted from the quantum blurring of the horizon position; while here, they result from
the string-scale blurring in the  transition time.
The resulting modified integral depends on an additional parameter $v_{shell}$,
\bea
{I}_{SC}(\omega'-\omega'';v_{shell})&=&\frac{1}{2\pi}\int\limits_{-\infty}^{\infty} dv\;  e^{i (v-v_{shell}) (\omega'-\omega'')} \;,
\eea
where $v_{shell}$ is to be regarded as a variable that is to be integrated against a normalized Gaussian probability distribution $P(v_{shell})$ with width $\sigma$. In the case of the frozen star, $\;\sigma\propto T_{trans}\;$.

The diagonal  and off-diagonal contributions  are then separated,
\bea
\langle {I}_{SC}(\omega'-\omega''; v_{shell}) \rangle &=& \delta(\omega'-\omega'')+ \left\langle\frac{1}{2\pi}\int\limits_{0}^{v_{shell}} dv^{\prime}\;  e^{i v^{\prime} (\omega'-\omega'')}\right\rangle, \label{ISC}
\eea
where $\;\langle f(v) \rangle=\int dv P(v) f(v)\;$.

Having mapped out a formally similar framework to that of \cite{slowleak}, we can now simply recall the end result. After an extremely long calculation, it was revealed that the number
operator for outgoing particles picks up an off-diagonal part~\footnote{It is explained in \cite{slowleak} as
to why the other two types of  two-point functions, that of  ingoing particles and mixed, are unaffected by the inclusion of quantum blurring.} that
 is a rather complicated combination of
 Gamma functions and exponential factors whose arguments contain the two out-frequencies
$\omega$, ${\widetilde \omega}$ and their difference $\;\Delta \omega = \omega-{\widetilde \omega}\;$.  Most important for current considerations is that the off-diagonal contribution scales with $\sigma/R$, which, on general grounds, should scale as $1/\sqrt{S_{BH}}$ \cite{density}.

Although the formal steps has been glossed over, it is useful to consider the actual mechanism that is responsible for the off-diagonal terms in the number operator. First, in the Hawking calculation, the number operator is measuring the overlap of two incoming waves at the same advanced time $v$. Such waves form an orthogonal basis in frequency-space, leading to adiagonal expression in frequency space  $\;I_C\sim \delta(\omega'-\omega)\;$. This  diagonal character is then preserved by the outgoing waves in the number operator after performing  a subsequent Fourier transform.  On the other hand, if these
modes become wavepackets that are smeared over a finite time scale, then the orthogonality condition  is modified in such a way
that the would-be delta function becomes a Gaussian and  the strength of the off-diagonal elements in the  number operator scales linearly with the duration of the smearing scale.
This happens in part because there are two subsequent Fourier transorms that are each weighted by a Gaussian. The first is the ``expectation value''
appearing in Eq.~(\ref{ISC}) and the second transforms a function of $\omega'-\omega''$  into  a function of $\;\Delta \omega = \omega-{\widetilde \omega}\;$, namely,  the number oeprator.

\begin{figure}
\centering
\includegraphics[width = .7\linewidth]{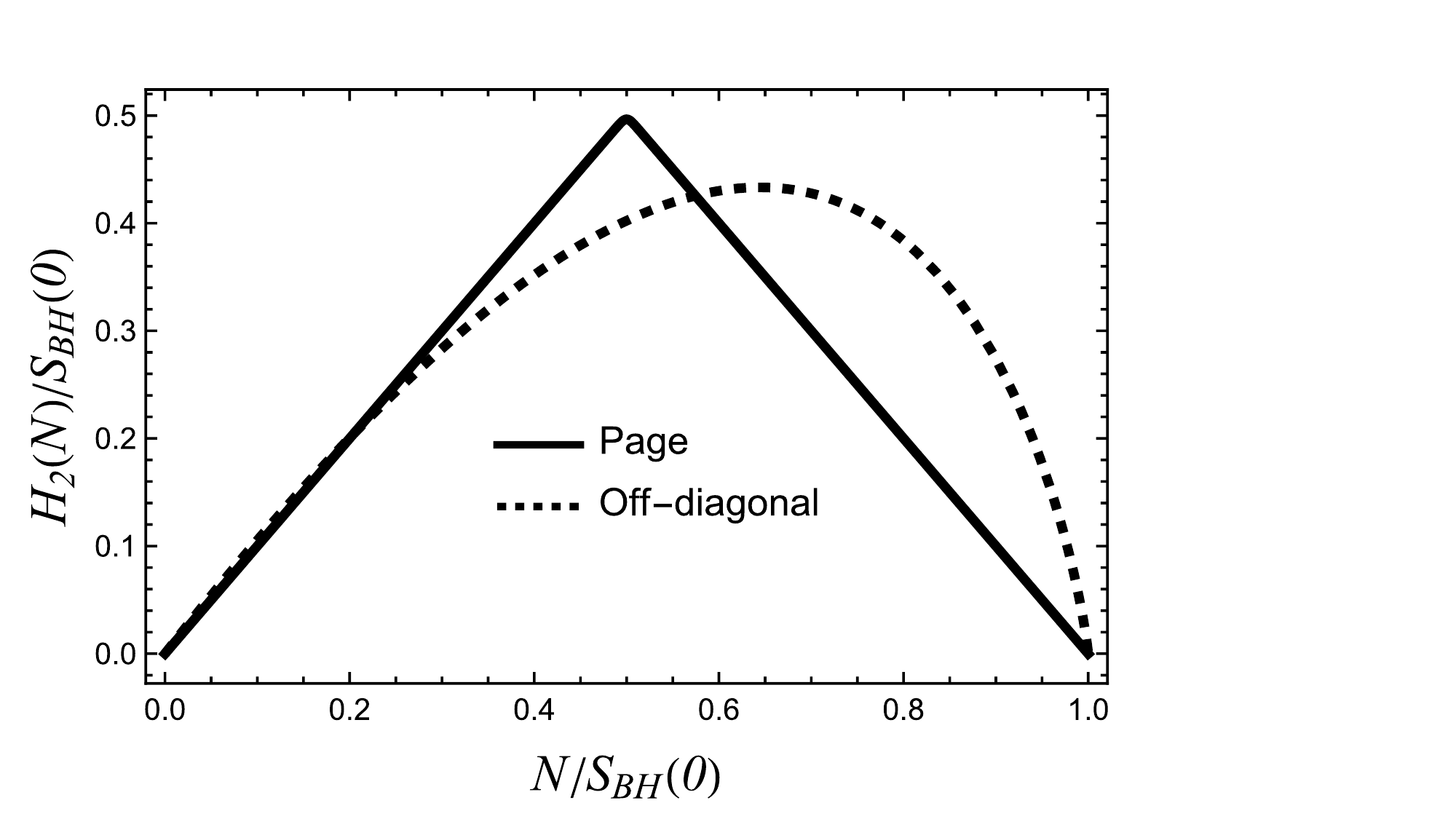}
\caption{Dependence of the R\'enyi entropy $H_2$ of the BH radiation on the  number of emitted particles $N$ for the case of off-diagonal elements discussed in the text (dashed). For comparison, the solid line depicts the prediction of the Page model. Figure reproduced from \cite{density}.
\label{fig:page}}
\end{figure}

Another long calculation, this time in
\cite{density}, reveals that an off-diagonal  correction of this strength in  the number operator  translates into off-diagonal elements  in the multi-particle density matrix
that are suppressed by a factor of $(\sigma/R)^2$ with respect to the diagonal elements. Consequently,
the off-diagonal contributions  are exponentially suppressed at early times but will become significant and eventually dominate, once the number of emitted particles $\;N\lesssim S_{BH}\;$ is enough to compensate for the suppression factor  $\;(\sigma/R)^2 \gtrsim 1/S_{BH}\;$.
This tipping point occurs near the midway point of evaporation, the so-called
Page time \cite{page}. Moreover, this type of relation between diagonal and off-diagonal elements is exactly what is needed to reproduce
the famous Page curve \cite{page}, which can be viewed as evidence of a unitary evaporation process. That this is so  has already been verified in \cite{density}. Here, we recall our previous form of the entropy-versus-time plot, reproduced in Fig.~\ref{fig:page} and refer back to the original source for mathematical details.

Tracking this flow of information over the lifetime of the star is operationally unfeasible. Hence, this form of density matrix  is in agreement with a previous argument that a standard BH and a BH mimicker, such as the frozen star, should be observationally  indistinguishable when in equilibrium  \cite{ridethewave}. Observational differences would rather  require an out-of-equilibrium event such as a BH merger in which the BH is emitting a macroscopic amount of energy in the form of gravitational waves.

\section{Conclusion}

Our main lesson is that the frozen star may be unique among  ultracompact but horizonless objects in that it is able to reproduce, up to perturbative corrections, the same entropy--area law as that of a conventional BH.  We have also shown that there is a natural choice for a Euclidean instanton which mediates the phase transition between the interior geometries  of an incipient BH  and a frozen star. Standard techniques were used to reveal  a probability of transition that goes as $e^{-S}$, just  as expected via statistical reasoning. Moreover, we have argued, following Mathur,  that the large degeneracy of the frozen star  predicts a total transition probability of order unity. Meaning that the transition from  a collapsing matter system to a frozen star would be commonplace if it is at all possible.

An unexpected consequence  of the probability-of-transition calculation is that it yields  an  extremely large
action for the intervening layer, being of the same order as the bulk action of the frozen star itself. The actual matter content, however, is suppressed by a factor of
$\lambda$, the shell width in Schwarzschild units, as a rather simple calculation reveals. A possible interpretation is that this matter might be in a highly excited state, which is suggestive of a BH ``firewall'' \cite{FW}.  If this  interpretation
of the large action  is indeed correct, it would be interesting to understand why a firewall-like structure is associated with a horizonless object.

It should be reemphasized that our calculation of the entropy is fundamentally geometric in that it is based directly on the Euclidean-action method. The energy density appears but has no obvious connection to matter fields in the context  of the frozen star model. The microscopic origin of the matter densities must rather be gleaned from the polymer model which is a string-theoretic description of the same compact object. The frozen star and collapsed polymer models are supposed to be two different perspectives of the same physical object. That the two models  are describing the same object  is supported by the similarities of this Euclidean geometry with that of the punctured cigar which is discussed in \cite{puncture}. In the punctured cigar, the Euclidean time circles  are decreasing with $r$ from the outer surface to the origin and are cut off, or ``punctured'', close to the $\;r=0\;$ tip. The frozen star geometry  puncture comes about because of its small, central, regularized zone. This similarity  is relevant  because the polymer model, once  Euclideanized, can also  be viewed as a condensate for the winding modes of the thermal scalar \cite{AW,HP,DV}. We hope to solidify this connection in the future, as BH physics --- which is becoming more and more accessible as gravitational-wave data collection and  analysis  improves --- could then provide a window into fundamental string theory.

\section*{Acknowledgments}
We thank Yoav Zigdon for discussions.
The research is supported by the German Research Foundation through a German-Israeli Project Cooperation (DIP) grant ``Holography and the Swampland'' and by VATAT (Israel planning and budgeting committee) grant for supporting theoretical high energy physics.
The research of AJMM received support from an NRF Evaluation and Rating Grant 119411 and a Rhodes  Discretionary Grant SD07/2022. AJMM thanks Ben Gurion University for their hospitality during his visit.

\end{document}